\documentclass{aa}
\usepackage{amssymb}
\usepackage{graphics}
\usepackage{natbib}
\usepackage{aas_macros}

\defcitealias{1995MNRAS.277..471S}{SPA95}
\defcitealias{1998ApJ...502..788T}{T98}
\defcitealias{1999A&A...341..842W}{WS99a}
\defcitealias{1999A&A...350..129W}{WS99b}
\defcitealias{2001A&A...366..840W}{WS01}
\defcitealias{1997MNRAS.291..651P}{PS97}

\bibpunct{(}{)}{;}{a}{}{,}
\newlength{\arraycolseporig}
\title{Tidal interaction of a rotating 1~$\mathrm{M}_\odot$ star with
  a binary companion}

\author{G.J. Savonije \&  M.G. Witte}

\institute{ Astronomical Institute `Anton Pannekoek', University of
  Amsterdam, Kruislaan 403, 1098 SJ Amsterdam} 
\date{received ?? ; accepted ??}

\begin{document}
\abstract{ We calculate the tidal interaction of a uniformly rotating
  1~$\mathrm{M}_\odot$ star with an orbiting companion at various
  phases of core hydrogen burning from the ZAMS to core hydrogen
  exhaustion.  By using the traditional approximation we reduce the
  solution of the non-adiabatic oscillation equations for the tidal
  forcing of a rotating star to a one dimensional problem by solving
  a separate eigenvalue problem for the angular dependence of the
  tidal perturbations.  The radial oscillation equations are then
  solved by using finite differencing on a fine grid with an implicit
  matrix inversion method like for stellar evolution calculations.  We
  are able to identify resonances with gravity \mbox{(g-)modes} and
  quasi-toroidal \mbox{(q-)modes} with up to $\simeq 1000$ radial nodes in
  the more evolved stellar models. The resulting tidal torque is
  calculated down to low forcing frequencies close to corotation.
  For low tidal frequencies we find significant response due to inertial
  \mbox{(i-)modes} in the convective envelope.  The inertial modes are damped
  by turbulent dissipation in the envelope and generate a relatively
  high torque-level in the low frequency region where the (retrograde) high 
  radial order g-mode resonances become tidally inefficient due to their rotational
  confinement to the stellar equator and strong damping by radiative losses.
  For still lower retrograde forcing   frequencies we find a large number 
  of closely spaced weakly damped quasi-toroidal q-mode resonances. 
  Our results indicate that effects   related to stellar rotation   
  can considerably enhance the speed of tidal evolution in low-mass binary systems.
  }
\maketitle

\section{Introduction}
Spectroscopic observations show there is a fairly well determined
orbital period below which main sequence binaries with solar-like
stars have essentially circularized orbits. This circularization
period appears to be $P_\mathrm{circ} \simeq 11$--$12$~days
\citep{1991A&A...248..485D,1992AJ....104..774L,1992bats.proc..278M}.  
In wider 
binary systems the tidal evolution is believed to be too slow to
achieve circularization during the main sequence phase of solar-type
stars. Since Zahn's (\citeyear{1977A&A....57..383Z}) review of tidal
effects there has recently been a renewed interest in tides in
solar-type binary system, because the circularization times
predicted by Zahn's analysis are too long to explain the
observations.  Zahn considered the interaction of turbulent eddies
with the hydrostatic \emph{equilibrium tide}, i.e. the low
frequency limit of the tidal response, as the dominant tidal mechanism
in low mass stars with convective envelopes. Although this mechanism
seems consistent with observed binary systems containing giant stars
\citep{1995A&A...296..709V}, it appears to fall short by orders of
magnitude when compared with observed solar-type binary systems
\citep[e.g.][]{1997A&A...318..187C}.  \citet{1997ApJ...486..403G} also
applied Zahn's formalism and concluded that it predicts a
circularization period of only $P_\mathrm{circ} \simeq 2.2$~days if
one allows for the reduced efficiency of the viscous damping when the
convective timescale of turbulent eddies is longer than the typical
tidal forcing period. They find $P_\mathrm{circ} \simeq 6$~days if one
ignores this effect.  Because the tidal torque scales as
$P_\mathrm{orb}^{-4}$, where $P_\mathrm{orb}$ is the orbital period, this
forms a significant discrepancy with the observationally inferred
$P_\mathrm{circ}\simeq 11$~days.

\citet[][from now on \citetalias{1998ApJ...502..788T}]{1998ApJ...502..788T} 
calculated the \emph{dynamical tide} raised by a close companion on a 
non-rotating solar-type star. 
For dynamical tides the interaction between the tidal
flow and the stellar normal modes of oscillation is taken into account
\citep{1941MNRAS.101..367C}.  The stratification in low mass main
sequence stars is complementary to that of massive stars and consists
of a radiative core surrounded by a convective envelope. Near the
boundary of the convective envelope the tides can excite g-modes in
the radiative core, where radiative damping is relatively weak.
Additional damping occurs in the convective envelope by the
interaction of the tidal flow with turbulent eddies.
\citetalias{1998ApJ...502..788T} allowed for damping due to the
interaction with these turbulent convective eddies by first order
perturbation analysis, while radiative damping was treated by a WKB
treatment of non-adiabatic terms.  They noted that the equilibrium
tide applies to forcing frequencies much smaller than the
Brunt-V\"{a}is\"{a}l\"{a} frequency $|\mathcal{N}^2|^{\frac{1}{2}}$
(see below) and that for solar-type stars with relatively small values
of $|\mathcal{N}^2|^{\frac{1}{2}}$ in the lower regions of the
convective envelope the convective timescale is too long for
essentially instantaneous adjustment to the tidal flow. Under these
circumstances the equilibrium tide is a rather poor approximation for
the tidal interaction and predicts a too strong effect.
\citetalias{1998ApJ...502..788T} found that for a fixed spectrum of
normal modes in a non-rotating star the tidal evolution of solar-type
binary systems is controlled essentially by the non-resonant dynamical
tide. To explain the observed $P_\mathrm{circ}$ would require a
viscosity that is $\simeq 50$ times larger than predicted by simple
mixing length estimates. \citet{1998ApJ...507..938G}, in their WKB
treatment of dynamical tides, considered non-linear damping of excited
g-modes near the stellar centre but they estimate this cannot explain
the discrepancy with the observations.

In \citetalias{1998ApJ...502..788T} the stellar normal mode
spectrum was considered fixed, i.e. for simplicity effects of stellar
rotation and stellar evolution were ignored. 
However, in orbits with a
significant eccentricity the tidal forcing is characterised not by one
but by a spectrum of forcing frequencies which can potentially excite
one or more of the normal modes in the binary stars.  It is often
argued that the chance of hitting a (narrow) resonance is small and
that --when it happens-- the enhanced tidal interaction rapidly moves
the system through the resonance without much orbital evolution.
This ignores the possibility of resonance locking. When
stellar rotation is taken into account, the tides can excite both
prograde (propagating in the direction of rotation) and retrograde
(counter to rotation) oscillation modes, depending on the stellar
rotation rate and orbital parameters. Excitation of prograde modes
gives rise to spin-up (positive torques), while retrograde tidal
forcing yields negative torques and spin-down.  Calculations for
massive binaries \citep[][from now on
\citetalias{1999A&A...341..842W,1999A&A...350..129W,2001A&A...366..840W}]
{1999A&A...341..842W,1999A&A...350..129W,2001A&A...366..840W} indicate
that often a weak high order prograde orbital harmonic is driven into
resonance with a stellar gravity \mbox{(g-)mode} by the simultaneous action
of a strong low frequency retrograde orbital harmonic which spins the
star down, increasing the relative forcing frequency of the high
orbital harmonic towards resonance. A similar driving effect can be
caused by stellar evolution induced spin-down.  The near-resonant weak
prograde orbital harmonic tend to spin the star up, counteracting the
retrograde driving ever more strongly when the peak of the resonance is
approached. The two counteracting (retrograde and prograde) orbital
harmonics then tend to lock each other at resonance with prolonged,
considerably enhanced, tidal interaction. The effectiveness of the locking 
process is sensitive to the magnitude of the orbital eccentricity and to the
low-frequency retrograde tidal forcing.  Unfortunately, the
low-frequency forcing excites high order (strongly damped) retrograde
g-modes with short wavelengths which requires a very large number
of meshpoints for finite difference calculations.  The 2D oscillation
code \citep{1997MNRAS.291..633S} which accounts for the full Coriolis
force is not suited for the low frequency calculations with low-mass
stars.

In this paper we calculate the fully non-adiabatic tidal interaction
with a uniformly rotating solar-type star by applying the traditional
approximation for which the $\vartheta$-component
of the stellar angular velocity is neglected. This is a reasonable
approximation for low oscillation frequencies which simplifies the
tidal problem significantly in that it allows a full separation of the
tidal perturbations in the three spherical coordinates, like for a
non-rotating star.  Here we present tidal torque calculations (down to
low forcing frequencies) for a uniformly rotating
1~$\mathrm{M}_\odot$ star at various stages of core hydrogen burning.

\section{Basic tidal oscillation equations}
We consider a uniformly rotating main sequence star with mass
$M_\mathrm{s}=1$~$\mathrm{M}_\odot$ and radius $R_\mathrm{s}$.  We assume the
star's angular velocity of rotation $\vec{\Omega}_\mathrm{s}$ to be
much smaller than its break-up speed, i.e.
$(\Omega_\mathrm{s}/\Omega_\mathrm{c})^2\ll 1$, with
$\Omega_\mathrm{c}^2=GM_\mathrm{s}/R_\mathrm{s}^3$, so that effects of
centrifugal distortion ($\propto \Omega_\mathrm{s}^2$) may be
neglected to a first approximation. We wish to study the response of
this uniformly rotating star to a perturbing time-dependent tidal
force. The Coriolis acceleration is proportional to
$\Omega_\mathrm{s}$ and we consider its effect on the tidally induced
motions in the star. We use nonrotating spherical coordinates
($r,\vartheta,\varphi)$, with the origin at the primary star's centre,
whereby $\vartheta=0 $ corresponds to its rotation axis which we
assume to be parallel to the orbital angular momentum vector.

Let us denote the displacement vector in the star by $\vec{\xi}$ and
perturbed Eulerian quantities like pressure $P'$, density $\rho'$,
temperature $T'$ and energy flux $\vec{F'}$ with a prime. The
linearised hydrodynamic equations governing the non-adiabatic response
of the uniformly rotating star to the perturbing potential
$\Phi_\mathrm{T}$ may then be written as 
\begin{eqnarray}
&&\left[ \left(\frac{\partial}{\partial t} + \Omega_\mathrm{s} \frac{\partial }
    {\partial\varphi}\right)
  v'_i\right] \vec{e}_i+ 2 \vec{\Omega}_\mathrm{s} \times \vec{v'}=
\nonumber\\
  &&\quad -\frac{1}{\rho}
  \nabla P' + \frac{\rho'}{\rho^2} \nabla P - \nabla \Phi_\mathrm{T},
  \label{eqmot}\label{eq:1}
\end{eqnarray}
\begin{equation}
  \left(\frac{\partial}{\partial t} + \Omega_\mathrm{s} 
    \frac{\partial }{\partial\varphi}\right) \rho' +
  \nabla\cdot\left(\rho \vec{v'} \right) =0, \label{eqcont}
\end{equation}
\begin{equation} 
  \left(\frac{\partial }{\partial t} + \Omega_\mathrm{s}
    \frac{\partial }{\partial \varphi}\right) \left[ S' +
    \vec{\xi}\cdot \nabla S \right]=-\frac{1}{\rho T} \nabla\cdot\vec{F'}, \label{eqe}
\end{equation}
\begin{equation}
  \frac{\vec{F'}}{F}=\left(\frac{\mathrm{d}T}{\mathrm{d}r}\right)^{-1} \left[ \left(\frac{3
        T'}{T} -\frac{\rho'}{\rho} -\frac{\kappa'}{\kappa} \right) \nabla
    T + \nabla T' \right] \label{eqf} 
\end{equation}
where $\kappa$ denotes the opacity of stellar material and $S$ its
specific entropy. These perturbation equations represent,
respectively, conservation of momentum, conservation of mass and
conservation of energy, while the last equation describes the
radiative diffusion of the perturbed energy flux. For simplicity we
adopt the \citet{1941MNRAS.101..367C} approximation, i.e. we neglect
perturbations to the gravitational potential caused by the primary's
distortion. We also neglect perturbations of the nuclear energy
sources and of convection.

Let us here consider the simplest case of a close binary system in
which the orbit is circular with angular velocity $\omega$ and orbital
separation $D$. The dominant tidal term of the perturbing potential is
then given by the real part of
\begin{equation}
  \Phi_\mathrm{T}(r,\vartheta,\varphi,t)= f r^2 \, P^2_2 (\mu) \,e^{ \mathrm{i} 
(\sigma  t -2 \varphi)} \label{eqpot}
\end{equation}
where $\sigma=2 \omega$ is the forcing frequency in the inertial frame,
$\mu=\cos\vartheta$, $P^2_2(\mu)$ is the associated Legendre polynomial 
and $ f= -\frac{G  M_\mathrm{p}}{4 D^3},$ with $M_\mathrm{p}$ the companion's mass.
After separating the $\varphi$-dependence, Eqs.~(\ref{eqmot}--\ref{eqf})
form a 2D problem in the ($r,\vartheta$) meridional plane of
the perturbed star. Let us now apply the traditional approximation in order to 
attain a considerable simplification of the tidal problem.

\section{The traditional approximation}
For $\Omega_\mathrm{s} \neq 0$ the solutions of Eqs.~(\ref{eqmot}--\ref{eqf})
 are no longer fully separable into $r$-,
$\vartheta$- and $\varphi$- factors due to the Coriolis force.
However, in the traditional approximation 
separability is retained by neglecting the $\vartheta$- component of
the angular velocity  \citep[e.g.][]{U89}. This is done because the radial 
motions are expected to be small in the stably stratified layers of the star,
especially for low-frequency high-order g-modes. For this reason, it
is thought to be a reasonable approximation for low oscillation
frequencies. Furthermore, a  spectrum of rotationally governed
inertial modes as well as (quasi-)toroidal modes is also present in
this approximation \citep{1995MNRAS.277..471S}.

When only the radial component of $\vec{\Omega}_\mathrm{s}$ is retained, the
$\vartheta$- and $\varphi$-components of the equation of motion can be
written: 
\begin{equation}
  -{\bar{\sigma}}^2\xi_{\vartheta} -2\mathrm{i}\Omega_\mathrm{s}{\bar{\sigma}}\cos\vartheta
  \xi_{\varphi}=
  -\frac{1}{r\rho}\frac{\partial P'}{\partial \vartheta} -\frac{1}{r}\frac{\partial
    \Phi_\mathrm{T}}{\partial \vartheta},\label{CRAP}
\end{equation}
\begin{equation}
  -\bar{\sigma}^2\xi_{\varphi}
  +2\mathrm{i}\Omega_\mathrm{s}\bar{\sigma}\cos\vartheta \xi_{\vartheta}=
  \frac{\mathrm{i}m}{r\rho\sin\vartheta } P'
  +\frac{\mathrm{i}m}{r\sin\vartheta} \Phi_\mathrm{T}.\label{CRAP1}
\end{equation}
We have expressed the velocity perturbations in terms of the
displacement vector by the relation $\vec{v'}=\mathrm{i} \bar{\sigma}
\vec{\xi}$, with $\bar{\sigma}=\sigma-m \Omega_\mathrm{s}$ the
oscillation frequency in the corotating frame, where $m=2$ in our
study. We assume $m$ to be always positive, whereby a retrograde oscillation 
(propagation direction counter to the sense of stellar rotation)
corresponds with negative oscillation frequencies $\bar{\sigma}$.

\subsection{Separation of variables}

A separation of variables can be performed by writing, see
 \citet{1997MNRAS.291..651P}:
\[ 
\xi_{\vartheta}=\sum_{n=1}^{\infty}\mathcal{F}_n(\vartheta)
D_n(r) \mbox{\hspace{1cm}}
\xi_{\varphi}=\sum_{n=1}^{\infty}\mathcal{G}_n(\vartheta) D_n(r)
\]
\begin{equation}
P' =\sum_{n=1}^{\infty}X_n(\vartheta)
  W_n(r),
\end{equation}
with $\xi_r$, $T'$ and $\rho'$ having expansions of the same form as that for
$P'$. Here $\mathcal{F}_n,\mathcal{G}_n$ and $X_n$ are functions of $\vartheta$ chosen
to obey the relations
\[ 
-\bar{\sigma}^2\mathcal{F}_n
-2\mathrm{i}\Omega_\mathrm{s}{\bar{\sigma}}\cos\vartheta
\mathcal{G}_n= -{\frac{\mathrm{d} X_n}{\mathrm{d}\vartheta}} 
\] 
and 
\[
-{\bar{\sigma}}^2\mathcal{G}_n
+2\mathrm{i}\Omega_\mathrm{s}\bar{\sigma}\cos\vartheta \mathcal{F}_n=
\frac{\mathrm{i}m}{\sin\vartheta} X_n.
\]

We obtain an equation for $X_n(\vartheta)$ by imposing the constraint
\[ 
\frac{1}{\sin\vartheta}\frac{\mathrm{d}\left( \sin\vartheta\mathcal{F}_n\right)}
 {\mathrm{d} \vartheta} -\mathrm{i} m\frac{\mathcal{G}_n}{\sin\vartheta}
=-\frac{\lambda_n}{{\bar{\sigma}}^2} X_n,
\] 
where $\lambda_n$ is a constant. Then $X_n$ must satisfy the second-order
equation obtained from
\begin{equation}
  \frac{1}{\sin\vartheta}\frac{\mathrm{d} \mathcal{Q}_n}{\mathrm{d}\vartheta} +
\frac{m x\cos\vartheta}{\sin^2\vartheta}\mathcal{Q}_n
=X_n\left(\frac{m^2}{\sin^2\vartheta}-\lambda_n\right)\label{lamb1}
\end{equation}
with
\begin{equation}
  \mathcal{Q}_n=\frac{\sin\vartheta}{(1-x^2\cos^2\vartheta)}\left( \frac{\mathrm{d} X_n}
      {\mathrm{d}\vartheta} -\frac{mx \cos\vartheta}{\sin\vartheta}X_n\right)\label{lamb2}
\end{equation}
whereby $X_n(\vartheta)$ is an eigenfunction with $\lambda_n$ the
associated eigenvalue. The solution depends on the rotation parameter
$x=2\Omega_\mathrm{s}/{\bar{\sigma}}$. The tidal problem we are interested in
corresponds with even solutions of $X_n(\mu)$ with $\mu=\cos{\vartheta}$ and 
$m=2$. 
For $\Omega_\mathrm{s}=0$ the functions $X_n(\vartheta)$ become the associated 
Legendre functions $P^2_{2n}(\cos\vartheta)$ with corresponding eigenvalues 
$\lambda_n=n~(n+~1).$ Normal modes of the rotating star correspond to normal
modes of the non-rotating star with $n~(n+~1)$ replaced by any
permissible $\lambda_n.$

Different $X_n(\vartheta)$ are orthogonal on integration with respect
to $\mu=\cos\vartheta$ over the interval $(-1,1)$.
If the perturbing potential is expanded in terms of the $X_n$ such
that
\begin{equation}
  \Phi_\mathrm{T}(r,\vartheta) =\sum_{n=1}^{\infty}\mathcal{V}_n(r) X_n(\vartheta), \label{potexp}
\end{equation} 
we find from~(\ref{CRAP}) and~(\ref{CRAP1}) that
\begin{equation}
  D_n(r)=\frac{1}{r\rho}W_n(r)+\frac{1}{r}\mathcal{V}_n(r).
\end{equation}
This equation is exactly the same as in the non-rotating case. The
same is true for the equation of continuity, except that
$n(n+1)$ is replaced by $\lambda_n$. If we ignore the energy equation (see below)
the tidal response will cnsist of a superposition of responses 
appropriate to non-rotating stars with $n~(n+1)$ replaced by $\lambda_n.$ 
Because the angular functions $X_n(\vartheta)$ are not spherical harmonics 
for non-zero rotation, all $n\equiv l$ are required in the superposition.

\subsection{The modified oscillation equations}
\label{sosceq}
Once we have solved the above eigenvalue problem for $\lambda_n$ the
linearised equations which describe the non-adiabatic forced tidal
oscillations for the $n$-th tidal component in expansion~(\ref{potexp})
become a 1-dimensional (radial) problem by factoring out the common
$\vartheta$-dependence factor $X_n(\vartheta)$ of all occurring
perturbations. The oscillation equations can now be expressed as
\begin{equation}
  \bar{\sigma}^2 \rho \xi_r-\frac{\mathrm{d}P'}{\mathrm{d}r}+\frac{\mathrm{d}P}{\mathrm{d}r} \left(\frac{\rho'}{\rho}\right)= 
  2\, \rho\, f\,  r, \label{eq1} 
\end{equation}
\begin{equation}
  \frac{1}{\rho r^2} \frac{\mathrm{d}\left(\rho r^2 \xi_r\right)}{\mathrm{d}r}= -\frac{\rho'}{\rho} + 
  \frac{\lambda_n}{\bar{\sigma}^2 r^2} \left[\frac{P}{\rho}\left(\frac{P'}{P}\right) + 
    f\, r^2\, \right],  \label{eq2}
\end{equation}
\begin{equation}
  \frac{P'}{P} - \Gamma_1 \frac{\rho'}{\rho} + \mathcal{A}\, \xi_r =
  \mathrm{i} \eta \left[ -\frac{1}{F r^2} \frac{\mathrm{d}(r^2 F'_r)}{\mathrm{d}r} +\lambda_n
    \frac{\Lambda}{r^2}\left(\frac{T'}{T}\right) \right], \label{eq3}
\end{equation}
\begin{equation}
  \frac{F'_r}{F}=\Lambda \frac{\mathrm{d}}{\mathrm{d}r} \left(\frac{T'}{T}\right)
  + (4 - \kappa_T) \frac{T'}{T} -(1 + \kappa_{\rho}) \frac{\rho'}{\rho}+
\frac{\kappa_X}{X} \frac{\mathrm{d} X}{\mathrm{d}r}  \xi_r, \label{eq4}
\end{equation}
where $\Gamma$'s represent Chandrasekhar's adiabatic coefficients, $X$ the hydrogen abundance,
\[
\Lambda= \left(\frac{\mathrm{d}\ln T}{\mathrm{d}r}\right)^{-1}, \mbox{\hspace{1.5cm}}
\mathcal{A}=\frac{\mathrm{d}\ln \rho}{\mathrm{d}r}- \frac{1}{\Gamma_1} \frac{\mathrm{d}\ln P}
{\mathrm{d}r},
\]
\[ 
\kappa_T=\left(\frac{\partial\ln \kappa}{\partial \ln T}\right)_{\rho},
\mbox{\hspace{.3cm}} \kappa_{\rho}=\left(\frac{\partial\ln \kappa}{\partial \ln
    \rho}\right)_T,
\mbox{\hspace{.3cm}} \kappa_{X}=\left(\frac{\partial\ln \kappa}{\partial \ln X}\right),
\]
\[ 
\eta=-(\Gamma_3-1) F/\left(\bar{\sigma} P\right), \mbox{\,\,\, with
  \hspace{0.3cm}} F=- \frac{4 a c T^3}{3 \kappa \rho} \frac{\mathrm{d}T}{\mathrm{d}r}
\]
the unperturbed radiative energy flux, and where the other constants
have their usual meaning.  The factor $\eta$ is a characteristic
radiative diffusion length. Note that the radiative diffusion
introduces a factor $\mathrm{i}$, so that the tidal perturbations are
complex-valued which expresses the induced phase-lags with respect to the
companion.

Although the energy equation is not separable, we have nevertheless simply replaced 
the $n(n+1)$ factor in the energy equation for a non-rotating star by $\lambda_n$ 
(eq. \ref{eq3}) to approximately take the horizontal radiative diffusion into account.
It appeared that the inclusion (exact for non-rotating stars) or non-inclusion  of this term  
yields almost the same values for the tidal torque, even when $|\lambda|$ becomes  
large at low retrograde forcing frequencies.

Finally, the oscillation equations are complemented by the linearised equation of state
\begin{equation}
  \frac{P'}{P}=\left(\frac{\partial\ln P}{\partial \ln T}\right) \frac{T'}{T}+ 
  \left(\frac{\partial\ln P}{\partial \ln \rho}\right) \frac{\rho'}{\rho}+
  \left(\frac{\partial\ln P}{\partial \ln \mu_a}\right) \frac{\mu'_a}{\mu_a} 
\end{equation} 
where $\mu_a$ is the mean atomic weight of the stellar gas. In
radiative layers we assume convection of oscillatory stellar material
without any diffusion, so that
\[ 
\frac{\mu'_a}{\mu_a}=-\frac{\mathrm{d}\ln \mu_a}{\mathrm{d}r}\,\, \xi_r.
\]

\subsection{Turbulent viscosity \label{svisc}}
The tidal oscillations are expected to be influenced by the turbulent
convective motions in the stellar envelope. We describe these
interactions in a heuristic way by means of a simple mixing length
approximation for turbulent viscosity. To this end we add in the
convective envelope a radial viscous force density
\[ 
f_{r,\mathrm{visc}} = 
\frac{\mathrm{i}\bar{\sigma}}{r^2} \frac{\partial}{\partial r}\left(\rho \nu r^2
  \frac{\partial\xi_r}{\partial r}\right)
\]
to the equation of motion (\ref{eq1}), with $\nu$ a turbulent viscosity coefficient.
In order to enable comparison we adopt the same form for the turbulent
viscosity coefficient as \citetalias{1998ApJ...502..788T}:
\begin{equation}
  \nu= \frac{L^2}{\tau_\mathrm{con}} 
  \left[ 1 +\left(\tau_\mathrm{con}\frac{\bar{\sigma}}{2 \pi}\right)^s
  \right]^{-1} \label{eqvisc} 
\end{equation}
where $L$ is the mixing length, taken to be twice the local pressure
scale height, $\tau_\mathrm{con}= |\mathcal{N}^2|^{-\frac{1}{2}}$ the
characteristic convective timescale with $\mathcal{N}^2=-g
\mathcal{A}$ the square of the Brunt-V\"{a}is\"{a}l\"{a} frequency,
whereby $g$ denotes the local acceleration of gravity.  The reduced
efficiency of the turbulent damping of high frequency oscillations by
the slow convective motions is taken approximately into account by the
factor in square brackets, where we adopt $s=2$. See
\citet{1977ApJ...211..934G}, \citet{1991ApJ...376..260G} and
\citet{1997ApJ...486..403G} for a discussion. The maximum value of the adopted 
viscosity coefficient $\nu$ is attained near the lower boundary of the convective 
envelope. In the stellar models used this maximum value ranges from a few times $10^{13}$  
to slightly more than $10^{15}$ in cgs units, whereby the latter value is reached for low 
forcing frequencies. 
Both the radiative and turbulent damping induce a phase lag of the tidal
response with respect to the forcing by the companion and that gives
rise to the tidal torques we are interested in. The torque-values both in between 
and near resonances are sensitive to the adopted turbulent viscosity law.

\subsection{The tidal torque}
The tidal torque integral corresponding with the $n$-th term of the
expansion~(\ref{potexp}) can be expressed as:
\begin{equation}
  \mathcal{T}_n(\bar{\sigma},\Omega_\mathrm{s})= \pi \zeta^2_n\,\, f \int_0^{R_\mathrm{s}} 
  \mathrm{Im}   \left[\rho'_n(r)\right] r^4 \mathrm{d}r \label{torq} 
\end{equation}
where $\pi$ results after the integration
over $\varphi$,   $f$ is the constant in the tidal potential defined by Eq.~(\ref{eqpot}), 
 $\mathrm{Im}$ stands for imaginary part, and

\begin{equation}
 \zeta^2_n=\frac{ \left[ \int_{-1}^1 P_2^2(\mu) X_n(\mu) \mathrm{d}\mu \right]^2} 
{\int_{-1}^1  X^2_n(\mu) \mathrm{d}\mu}.
\label{zeta_n}
\end{equation}
$\zeta_n$ is an overlap integral of the 
eigenfunctions $X_n(\mu)$ with $P^2_2(\mu)$, the associated Legendre
polynomial describing the $\mu$ variation of the dominant $l=m=2$ tide. 
The tidal torque is proportional to $\zeta_n^2$, which
for $n=1$ and $\Omega_\mathrm{s} \rightarrow 0$ attains the value 48/5.

The tidal torque follows by multiplying $\mathcal{T}_n$ by the factor $m=2$.  
Once we have, for a given forcing frequency $\bar{\sigma}$ and stellar angular
rotation speed $\Omega_\mathrm{s}$, first solved the $\lambda_n(x)$
eigenvalue problem and subsequently the radial oscillation Eqs.~(\ref{eq1}--\ref{eq4}),
the resulting $\rho'_n(r)$ can be substituted in the above integral to evaluate 
the tidal torque.

\section{Solution method}
\subsection{Determination of angular eigenvalue $\lambda_n(x)$}
For a given stellar rotation frequency $\Omega_\mathrm{s}$ and forcing
frequency $\bar{\sigma}$ the eigenvalues $\lambda_n$ can be determined
by numerically integrating the two ordinary differential Eqs.~(\ref{lamb1}) and~(\ref{lamb2}).
We have used a shooting method with fourth order Runge-Kutta
integration with variable stepsize \citep[e.g.][]{1992nrfa.book.....P}.  
To this end we rewrite the equations as
\begin{equation}
  \frac{\mathrm{d}X_n}{\mathrm{d}\mu}=-\frac{m x\, \mu}{1-\mu^2}\,X_n -\left(\frac{1-x^2 \mu^2}
    {1-\mu^2} \right) \,\mathcal{Q}_n, \label{Lamb1} 
\end{equation}
\begin{equation}
  \frac{\mathrm{d}\mathcal{Q}_n}{\mathrm{d}\mu}=-\left(\frac{m^2}{1 -\mu^2} -\lambda_n\right)\, 
X_n +
  \frac{m x \mu}{1 -\mu^2} \,\mathcal{Q}_n \label{Lamb2} 
\end{equation}
where the factor $x=2 \Omega_\mathrm{s}/\bar{\sigma}$ expresses the
importance of stellar rotation.

To enable integration away from $\mu=1$ it is convenient to write $
X_n(\mu)=(1 -\mu^2)^{\frac{m}{2}} Y_n(\mu) $ and expand $Y_n(\mu)$ in
a power series $ Y_n=c_0+c_1\, (\mu-1)+ c_2\, (\mu-1)^2+\ldots $ which
can be substituted in Eqs.~(\ref{Lamb1}) and~(\ref{Lamb2}) to determine
the coefficients $c_k$. We use $c_0$ as a free scaling parameter and
express coefficients $c_1$ to $c_4$ in terms of $c_0$, $x$ and 
the unknown eigenvalue $\lambda_n$.
$\mathcal{Q}_n$ can then be expressed in terms of $Y_n$ as
\[ 
\mathcal{Q}_n=\frac{(1-\mu^2)^{\frac{m}{2}}} {1-x^2 \mu^2} \, \left[ m
  \mu (1-x)\, Y_n-(1-\mu^2)\, \frac{\mathrm{d}Y_n}{\mathrm{d}\mu} \right].
\] 
For the even (in $\mu$) $m=2$ solutions the boundary conditions 
are $X_n=0$ at $\mu=1$ and $\mathcal{Q}_n=0$ at $\mu=0$.
We can now integrate equations (\ref{Lamb1})-(\ref{Lamb2}) from 
$\mu=1-\epsilon$ (with $\epsilon=10^{-4}$) to
$\mu=0$ with an estimated value for the eigenvalue $\lambda_n$. We
iterate by adjusting $\lambda_n$ until the integrated value for
$\mathcal{Q}_n$ is sufficiently close to zero for $\mu=0$.

\subsection{Numerical solution of radial oscillation equations \label{snumsol}}
Once $\lambda_n(x)$ is determined by means of the above numerical
method the radial oscillation Eqs.~(\ref{eq1}--\ref{eq4}) can be
solved for the same value of $\bar{\sigma}$ after prescribing the
usual boundary conditions  $\xi_r=F'=0$ at the stellar centre and by requiring
that the Lagrangian perturbations at the stellar surface obey:
$\delta P'=0$ and $\delta F'/F=4\, \delta T'/T$ (Stefan's law).
By transforming the differential equations into algebraic equations 
by means of finite differences on a spatial
mesh and applying matrix inversion similar to standard Henyey schemes
for stellar evolution  the radial equations 
can be solved \citep[e.g.][]{1983MNRAS.203..581S}. The viscous
diffusion term requires three-level finite differencing. It appears
that, because of the short wavelength of the tidal response for
low forcing frequencies $\bar{\sigma}$ one needs a large number of
meshpoints for a solar-type star: we have used 5000 radial points. The
2D oscillation code \citep{1997MNRAS.291..633S} which takes the
Coriolis force fully into account would require unrealistically long
integration times for such a fine radial grid in combination with an
adequate resolution in $\vartheta$. In order to obtain consistent
results for the resonances  we have used the traditional 
approximation also outside the low frequency inertial range
(although in Figure~2 we only show results for $|x|<1$).

\section{Stellar models}
\begin{figure}[htbp]
  \includegraphics{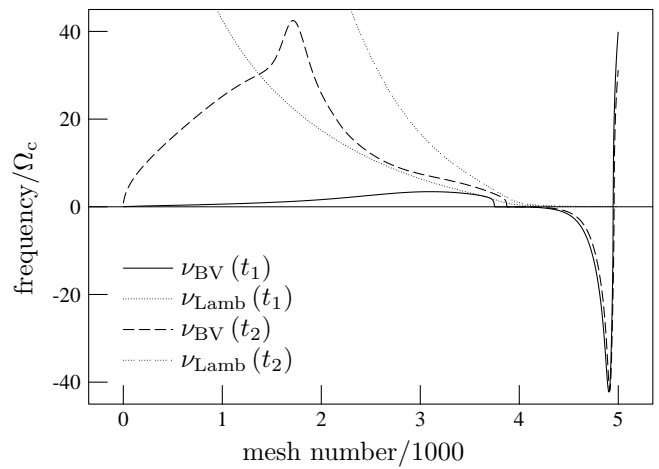}
  \caption{The modified Brunt-V\"{a}is\"{a}l\"{a} frequency 
    $\nu_\mathrm{BV}=\,\mathrm{sign}\,\mathcal{N} \sqrt{|\mathcal{N}|}$
    and Lamb-frequency in units of the stellar break-up speed
    $\Omega_\mathrm{c}$ versus meshpoint number for ($t_1:0.15$~Gyr)
    an unevolved 1~$\mathrm{M}_\odot$ model, and ($t_2:11$~Gyr) for a
    stellar model with central hydrogen abundance $X_\mathrm{c}\simeq
    0$.}
  \label{fBV} 
\end{figure}

We constructed the unperturbed stellar models for the solar-type star
with a recent version \citep{1995MNRAS.274..964P} of the stellar
evolution code developed by \citet{1972MNRAS.156..361E}.  The models
represent a spherical main sequence star of 1~$\mathrm{M}_\odot$ with
chemical composition given by various values of the central hydrogen
abundance $X_\mathrm{c}$ and $Z=0.02$.  The model was constructed by
using the OPAL opacities \citep{1996ApJ...464..943I}.  Fig.~\ref{fBV}
shows a frequency related to the Brunt-V\"{a}is\"{a}l\"{a}
frequency $\mathcal{N}$, namely $\nu_\mathrm{BV}=\,\mathrm{sign}\,\mathcal{N}
\sqrt{|\mathcal{N}|}$ in units of the stellar break-up speed
$\Omega_\mathrm{c}$ versus mesh point number for both an unevolved
stellar model and for a model near the end of core hydrogen burning.
In the latter model the square of the Brunt-V\"{a}is\"{a}l\"{a}
frequency has become quite large in the contracted helium core. The
radiative core, where the tidal oscillations occur, is made to contain
roughly 4000 meshpoints in order to resolve the short wavelength
response for low forcing frequencies.

\section{Results}

\subsection{The solution for the eigenvalues $\lambda_n$}
We only show the solution of the eigenvalue problem in the the low-frequency 
inertial regime, for which the absolute value of the rotation parameter 
$x=2\Omega_\mathrm{s}/\bar{\sigma}$ is larger than unity. 
Negative $x$-values correspond to retrograde, positive values to prograde forcing.

\subsubsection{g-mode solutions}
\begin{figure*}[htbp]
  \includegraphics{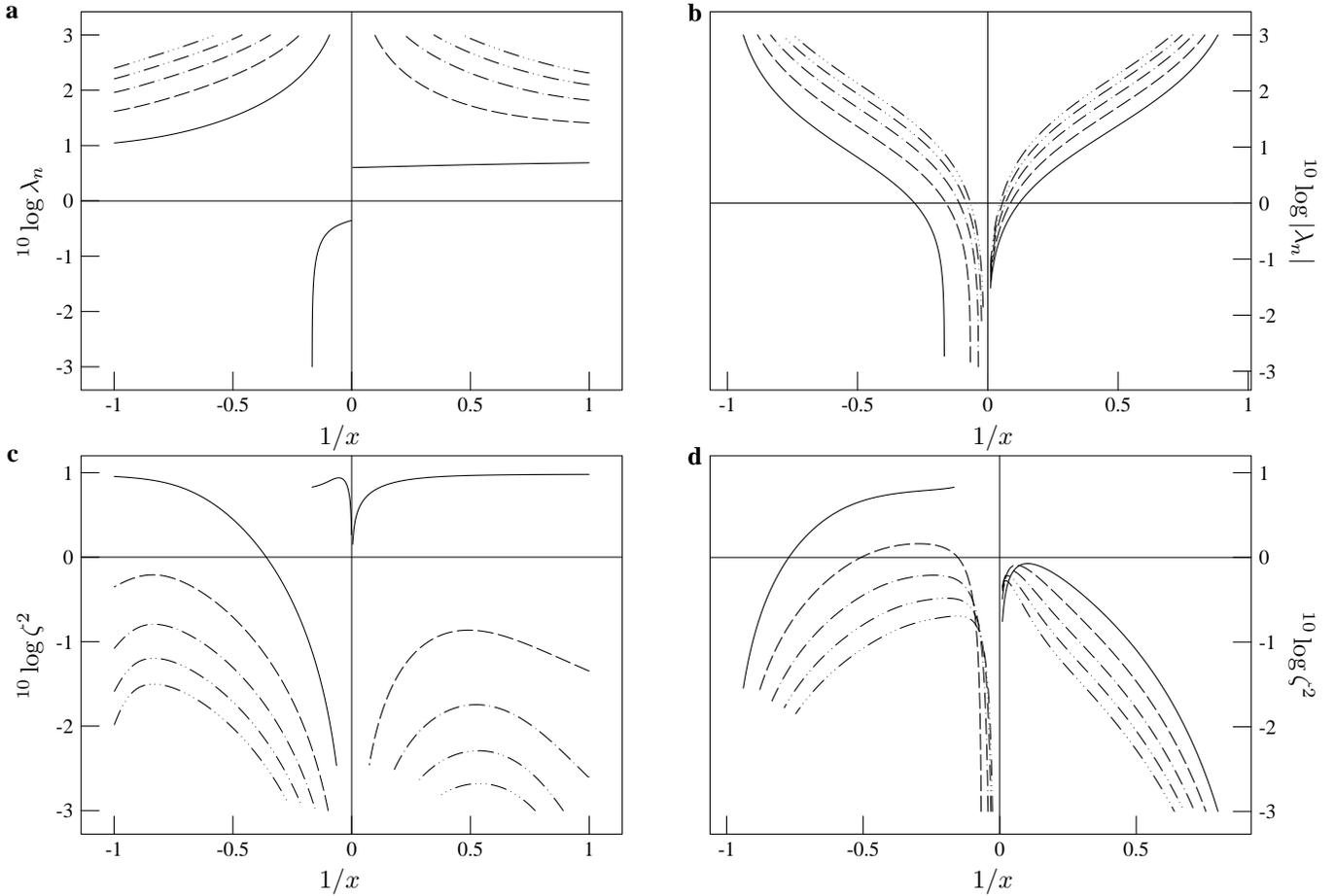}
  \caption{\textbf{a} The calculated positive eigenvalues $\lambda_n$ (for $n=1$ to 5) versus 
    $1/x=\bar{\sigma}/(2 \Omega_\mathrm{s})$ for m=2, corresponding to eigenfunctions with 
    even symmetry in $\mu$.     The full curves
    correspond to $n=1$, the uppermost curves to $n=5$. \textbf{b} The same
    as panel a but now for the negative eigenvalues. \textbf{c} The square
    of the tidal angular overlap integral $\zeta_n$ for $\lambda_n>0$ versus $1/x$
    with same line coding for n=1 to 5. \textbf{d} The square of 
    $\zeta_n$ for $\lambda_n<0$ versus $1/x$ with
    same line coding for n=1 to 5.}
  \label{feigenval} 
\end{figure*}

The calculated positive eigenvalues $\lambda_n$, with $n=1$ to 5,   
are  shown as a function of $1/x$  in Fig.~\ref{feigenval}a. 
For large values of $|1/x|$ the eigenvalues $\lambda_n$ approach the values 
$l(l+1)$ (with l=2n), corresponding to vanishing rotation. The values for 
$\lambda_1$ are consistent with those 
calculated in \citet{1997MNRAS.291..651P}. 
It can be seen that rotational effects become strong when 
$|1/x|\rightarrow 0$, i.e. when the oscillation frequency becomes 
small compared to the angular speed of the star.  

The large values for $\lambda_n$  for $|x|^{-1}\rightarrow 0$ are associated 
with eigenfunctions $X_n$ which attain small values for $1~ \ge~ \mu>>~ |x|^{-1}$, 
while the peak(s)  shift(s) to  $\mu=0$. This corresponds to the 
confinement of g-modes towards the stellar equator due to strong 
Coriolis effects near the poles for small oscillation frequencies.  
For the lowest eigenvalue this effect is strongest for the 
retrograde g-modes. Fig.~\ref{feigenval}c shows the angular overlap 
integral (\ref{zeta_n})
$\mathrm{^{10}log[\zeta^2_n]}$ as a function of $1/x$ for $\lambda_n>0$.  
Due to the equatorial confinement of the g-modes their coupling to the
tides becomes weak when $|1/x|$ becomes small, with strongly reduced
values of $\zeta_n$. The latter effect is enhanced due to the fact that the 
retrograde solutions have one node more in the $\vartheta$ direction than the 
corresponding prograde solutions. 

\subsubsection{r and q-mode solutions}
For $-1/6<x^{-1}<0$ there are solutions with small eigenvalues, see
Fig.~\ref{feigenval}a.
The solution with $\lambda_n$= 0 is a purely toroidal
oscillation with frequency corresponding to $x=-6$.  
This solution corresponds to the class of r-modes 
\citep{1978MNRAS.182..423P} with frequencies
\[ \bar{\sigma}_\mathrm{r}=-\frac{2 m \Omega_\mathrm{s}}{l(l+1)}.\]
For the lowest order $l=3$ (which couples to the dominant $l=m=2$
tide) this frequency coincides with $1/x=-1/6$. In previous papers
\citepalias{1999A&A...341..842W,1999A&A...350..129W,2001A&A...366..840W}
we loosely called the quasi-toroidal modes (with associated small but 
non-zero values of $\lambda_n$) also r-modes, 
but here (see section \ref{sgqmodes}) we have identified these modes up to
 high radial order for which the non-toroidal character becomes
significant and we denote them as \emph{q-modes}. Like g-modes, 
the peak(s) of the eigenfunctions for q-modes shift to small values
of $\mu$ when $x^{-1} \rightarrow 0$, although the equatorial confinement 
is less pronounced. It can be seen in Fig.~\ref{feigenval}c that the q-modes 
couple effectively to the $l=m=2$ tide, with the angular overlap integral 
 $\zeta^2 \simeq 10$. 

\subsubsection{i-mode solutions}
Solutions with $\lambda_n<0$ are associated with rotationally governed 
\emph{inertial} i-modes which can propagate in the approximately adiabatic 
convective envelope. 
Fig.~\ref{feigenval}b shows the absolute value of the eigenvalues associated
with inertial solutions for $m=2$ and with the adopted even symmetry in $\mu$
for the eigenfunctions $X_n$.
When $|x|^{-1} \rightarrow 1$ the eigenfunctions of the i-solutions  become more 
and more confined to the stellar poles, i.e. the coupling with the tide decreases 
when the effect of rotation becomes weaker (see Fig.~\ref{feigenval}d).
The eigenvalues for the prograde i-solutions vanish when $x^{-1} \rightarrow 0$, while
the eigenvalues for retrograde i-solutions tend to zero for small, but finite values 
of $|x|^{-1}$.  The eigenvalue corresponding to the fundamental inertial solution drops 
in fact to zero for $1/x \rightarrow -1/6$. At this point the fundamental inertial 
solution connects to the toroidal r-solution. The prograde solutions have one 
node more in the $\vartheta$ direction than the corresponding retrograde solutions.

\subsection{The eigenfrequencies: effects of stellar evolution}
\begin{figure*}[htbp]
  \includegraphics{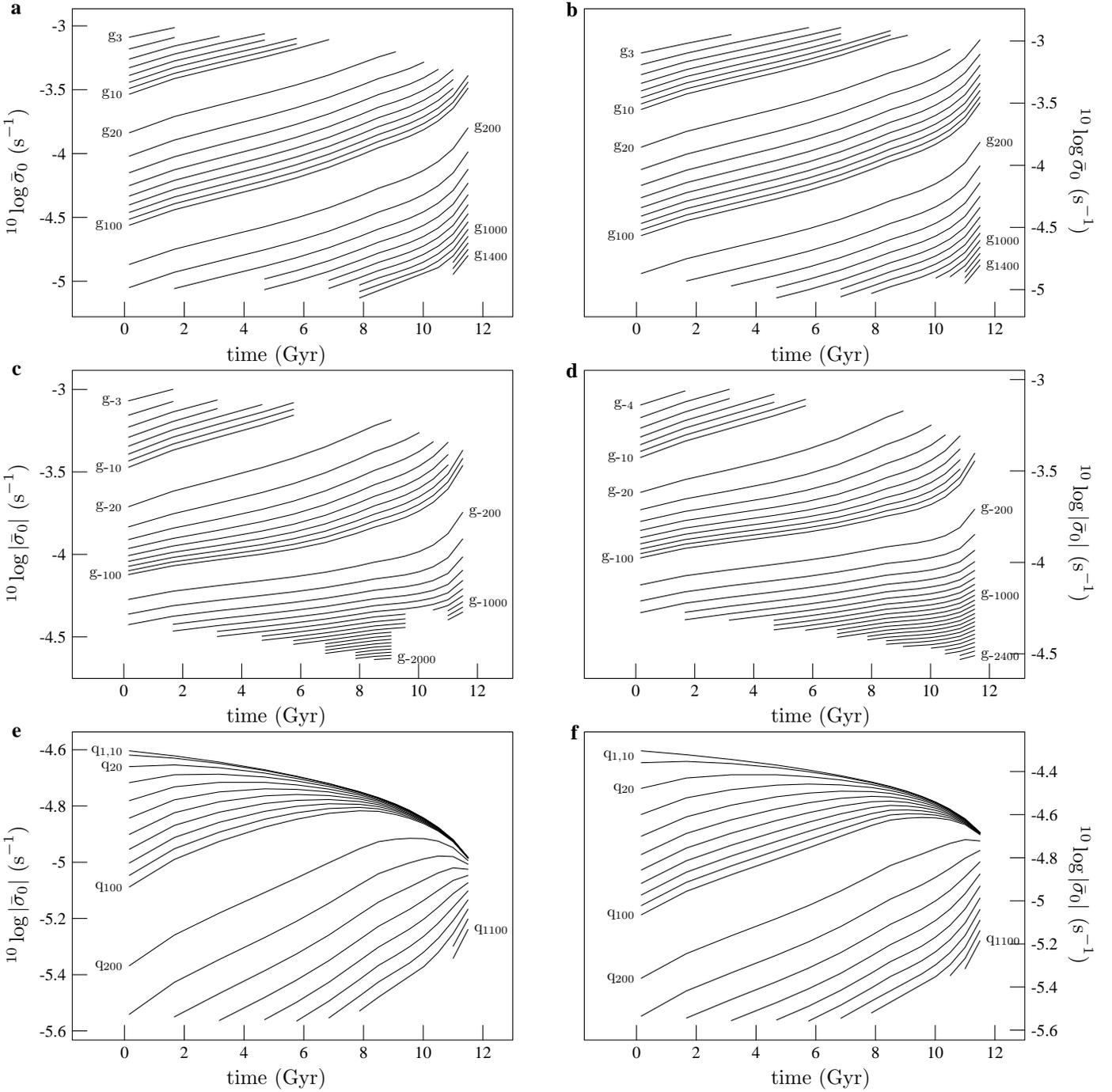}
  \caption{The variation of resonance frequencies $\sigma_{\mathrm{0}}$ 
           corresponding to $\lambda_1>0$ with stellar evolution. 
    \textbf{a} and \textbf{b} The prograde g-modes for rotation rate
    $\Omega_\mathrm{s}=0.1$ and $0.2$, respectively.  \textbf{c} and
    \textbf{d} The retrograde g-modes for rotation rate $\Omega_\mathrm{s}=0.1$ and
    $0.2$.  \textbf{e} and \textbf{f} The quasi-toroidal q-modes for rotation rate
    $\Omega_\mathrm{s}=0.1$ and $0.2$. }
  \label{fevol} 
\end{figure*}

By substituting the calculated values for the angular eigenvalue
$\lambda_1$ (for the given value of $x=2
\Omega_\mathrm{s}/\bar{\sigma}$) in Eqs.~(\ref{eq1})--(\ref{eq4})
 and solving these equations numerically for
a range of forcing frequencies $\bar{\sigma}$, the resonances with
stellar oscillation modes are readily found. By doing this for
several stellar input models, at various phases of core hydrogen
burning, one can follow the evolution of the normal mode resonances
of the 1~$\mathrm{M}_\odot$ star from the ZAMS to the end of core
hydrogen burning. We have calculated the resonances for a
non-rotating star, as well as for a star rotating at
$\Omega_\mathrm{s}=0.1$ and $0.2$ (in units of $\Omega_\mathrm{c}$).
Figs.~\ref{fevol}a--\ref{fevol}d show
the evolution of the resonance frequencies for prograde and
retrograde g-modes for stellar rotation speeds
$\Omega_\mathrm{s}=0.1$ and $0.2$, as the 1~$\mathrm{M}_\odot$ star
evolves away from the ZAMS. In both cases the g-mode frequencies
increase with age, as is expected from the fact that the stellar
core contracts, so that the acceleration of gravity increases in the
propagation region of the modes. We also find that the retrograde
g-modes have higher resonance frequencies (in absolute value) than
their prograde counterparts.

Figs.~\ref{fevol}e and \ref{fevol}f show the more complex evolution
of the quasi-toroidal q-modes when the star evolves away from the
ZAMS.  The frequency
of toroidal modes is proportional to the the rotation speed
$\Omega_\mathrm{s}$. When the star evolves away from the ZAMS its
radius increases, at first slowly, but near the end of core hydrogen
burning more quickly. Rotation at fixed
$\Omega_\mathrm{s}/\Omega_\mathrm{c}$ thus means that the stellar
rotation rate assumed for this figure slows in fact down with age
because our frequency unit $\Omega_c$ decreases when the star
expands. This tends to bring down the mode frequencies with age, as
is indeed the case for the low radial order q-modes. However, the
higher radial order q-modes correspond with larger eigenvalues
$\lambda_1$ and thus with an increasing non-toroidal character as
the radial node number $k$ rises, for which buoyancy effects become
important. This explains why the frequency of the high order q-mode
resonances increase with age, like g-modes. Comparing
Figs.~\ref{fevol}e and \ref{fevol}f shows that the resonance
frequencies of low-order q-modes do indeed scale with
$\Omega_\mathrm{s}$, and less so for the high order q-modes.

\subsection{Determination of the tidal torque}
Although the expansion (\ref{potexp}) should be made over all $n$
values, the strength of the tidal coupling decreases (Figs.~\ref{feigenval}c-d)
rapidly for higher orders, so that we will from now on consider only the first
eigenvalue $\lambda_1$, and ignore the intrinsically weaker coupling
of the higher order eigenfunctions.

For a given forcing frequency $\bar{\sigma}$ and stellar spin rate
$\Omega_\mathrm{s}$ we first determine $\lambda_1(x)$ and substitute
this in the set of Eqs.~(\ref{eq1}--\ref{eq4}).  After solving these
equations (section (\ref{snumsol}) and substituting the obtained
imaginary part of $\rho'$ in the torque integral~(\ref{torq})
the tidal torque can be evaluated. The calculations are done for a
perturbing companion of 1~$\mathrm{M}_\odot$ in \emph{circular}
orbit with binary separation $D=4 R_\mathrm{s}$, i.e. for a
\emph{fixed} ratio of $R_\mathrm{s}/D$. The values for the torque
integral $\mathcal{T}_1$ have to be multiplied by a factor $m=2$ to 
get the tidal torque $T$.

The tidal torque on a non-synchronously rotating solar-type star in
a given binary system with a companion of mass $M_p$ in a circular
orbit with arbitrary orbital separation $D$ is then found from
\[ 
T(\bar{\sigma}) =m \left(\frac{M_p}{\mathrm{M}_\odot}\right)^2 \,
\left(\frac{4 R_\mathrm{s}}{D} \right)^6\, \mathcal{T}_1
(\bar{\sigma})
\] 
where, for a stellar model with $X_c \simeq 0.4$ and a given value of
$\bar{\sigma}=\sigma -m \Omega_\mathrm{s}$ with $m=2$, $\mathcal{T}_1$ can be read from
Figs.~\ref{ftork1} to \ref{ftork3}.

\subsection{Comparison with \citetalias{1998ApJ...502..788T}'s calculation}
To check our results we have compared some of our torque-values for
the non-rotating case with those of
\citetalias{1998ApJ...502..788T}, as there exist no detailed results
for rotating stars to compare with. 
\setlength{\arraycolsep}{\arraycolseporig}
\begin{table}[htbp]
  \caption{Comparison of calculated torque values: the upper values are from
    \citetalias{1998ApJ...502..788T}, while the lower 
    values correspond to the calculations presented here.
    $T_\mathrm{peak}$ and $T_\mathrm{off}$ are the torque 
    values at the resonance peaks considered, and for neighbouring off-resonant frequencies, 
    respectively. The middle column gives the relative FWHM of the resonance peaks. \label{table1}}
  \[
  \begin{array}{||c|l|l|l||}  \hline 
    \rule[-2ex]{0ex}{5ex}
    P\mbox{ (d)}  &  T_\mathrm{peak}\mbox{ (cgs)} & \Delta \sigma/\sigma
    & T_\mathrm{off}\mbox{ (cgs)} \\
    \hline 
    \hline
    \rule{0ex}{3ex}
    4.2  &  3.2\times 10^{38}  & 3 \times 10^{-6}  &  4 \times 10^{35}\\
    &  3.4\times 10^{38}  & 7 \times 10^{-6}  &  7 \times 10^{35}\\
    \hline
    \rule{0ex}{3ex}
    8.5  &  2.8\times 10^{35}  & 4 \times 10^{-5}  &  6 \times 10^{34}\\
    &  3.6\times 10^{35}  & 4 \times 10^{-5}  &  5 \times 10^{34}\\
    \hline
    \rule{0ex}{3ex}
    12.4  &  6.3\times 10^{33}  & 2 \times 10^{-4}  &  5 \times 10^{33}\\
    &  1.5\times 10^{34}  & 2 \times 10^{-4}  &  7 \times 10^{33}\\
    \hline
  \end{array}
  \]
\end{table}
\setlength{\arraycolsep}{0pt}
 Table~\ref{table1} gives some
results of both calculations for three different circular orbits
with periods of $4.2$, $8.5$ and $12.4$~days. In the table we have scaled
our results to those of \citetalias{1998ApJ...502..788T}, i.e. we
use their normalisation of the tidal torques as in their Fig.~2. After comparing 
their torque values with our results  for non-rotating stars  we conclude that 
the two calculations yield quite similar results.

\subsection{Torque spectra for $\lambda_1>0$ (g- and q-modes) \label{sgqmodes}}
\begin{figure*}[htbp]
  \includegraphics{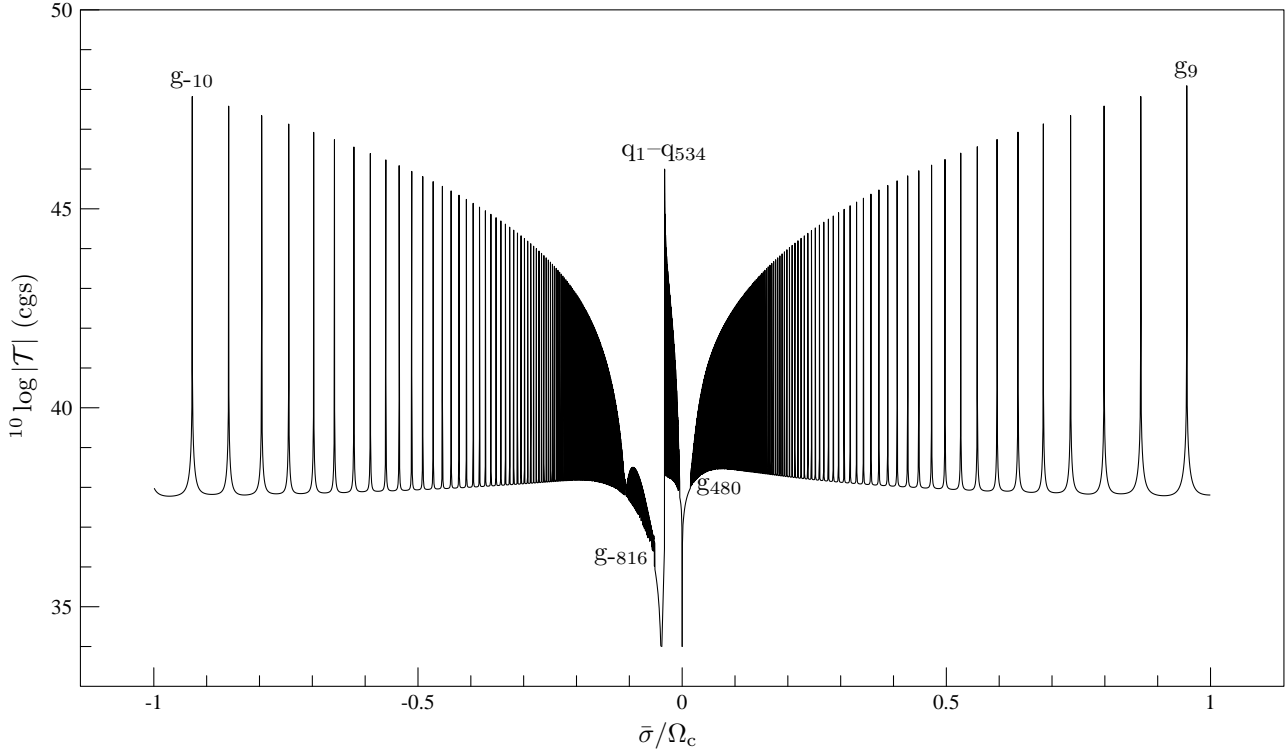}
  \caption{The torque integral $\mathcal{T}$  versus forcing frequency 
    $\bar{\sigma}$ for  $\lambda_1>0$.
    The calculation  is for a 1~$\mathrm{M}_\odot$ star with central hydrogen abundance
    $X_\mathrm{c}=0.4$, rotating at speed $\Omega_\mathrm{s}=0.1\,
    \Omega_\mathrm{c}$. The narrow dip before each (low order) resonance peak has
    been ignored in this plot. Negative values of $\bar{\sigma}$
    correspond to retrograde forcing, for which the sign of the
    torque should be read as negative.  Results are for a fixed
    ratio of orbital separation to radius $D/R_\mathrm{s}=4$ and
    companion mass $M_p=1$~$\mathrm{M}_\odot$. Many strong
    resonances with quasi-toroidal q-modes, shown here as an unresolved
    black peak just left of $\bar{\sigma}=0$, can be identified
    (with up to $\simeq 500$ radial nodes) for this stellar model. }
  \label{ftork1} 
\end{figure*}

After repeating the torque calculation procedure for a wide range of
frequencies ($|\bar{\sigma}| \le 1$, where all frequencies stated
without units are in units of the critical rotation rate
$\Omega_c$), we end up with a number of torque spectra, like the one
shown in Fig.~\ref{ftork1} for $\Omega_\mathrm{s}=0.1$ and
$X_\mathrm{c}=0.4$. Note that the actual height of the resonances depends on
the adopted turbulent viscosity law for the convective regions and  
on the amount of turbulent overshooting. We have assumed a Gaussian decay 
of the viscosity coefficient beyond the convection boundaries with a decay length 
of 0.5 pressure scale height.

The high frequencies near $|\bar{\sigma}|=1$ will never
occur in normal binaries with solar-type stars, but part of the
higher frequencies may be relevant to extreme systems with planets
in very  eccentric orbits. For $|\bar{\sigma}|<0.25$
the resonance spectrum of prograde and retrograde high radial order
g-modes becomes so dense that in Fig.~\ref{ftork1} it can no longer 
be resolved.  For decreasing forcing frequency $|\bar{\sigma}|$ the
radial wavelength of the g-modes decreases, so that the resonances 
fade away by severe radiative damping.  Also the tidal coupling strength 
$\zeta_1$ of (especially the retrograde) g-modes declines with $|\bar{\sigma}|$ 
(see Fig.~\ref{feigenval}c).  For the high frequency g-mode resonances  
turbulent viscosity in the (lower part) of the convective envelope provides 
the dominant damping. The turnover to predominant radiative damping occurs 
near $|\bar{\sigma}|\simeq 0.1$

We have zoomed in to determine the resonances with the stellar normal modes 
down to the (somewhat arbitrary) frequency where 
the resonances become too weak to trace down individually. 
\begin{figure*}[htbp]
  \includegraphics{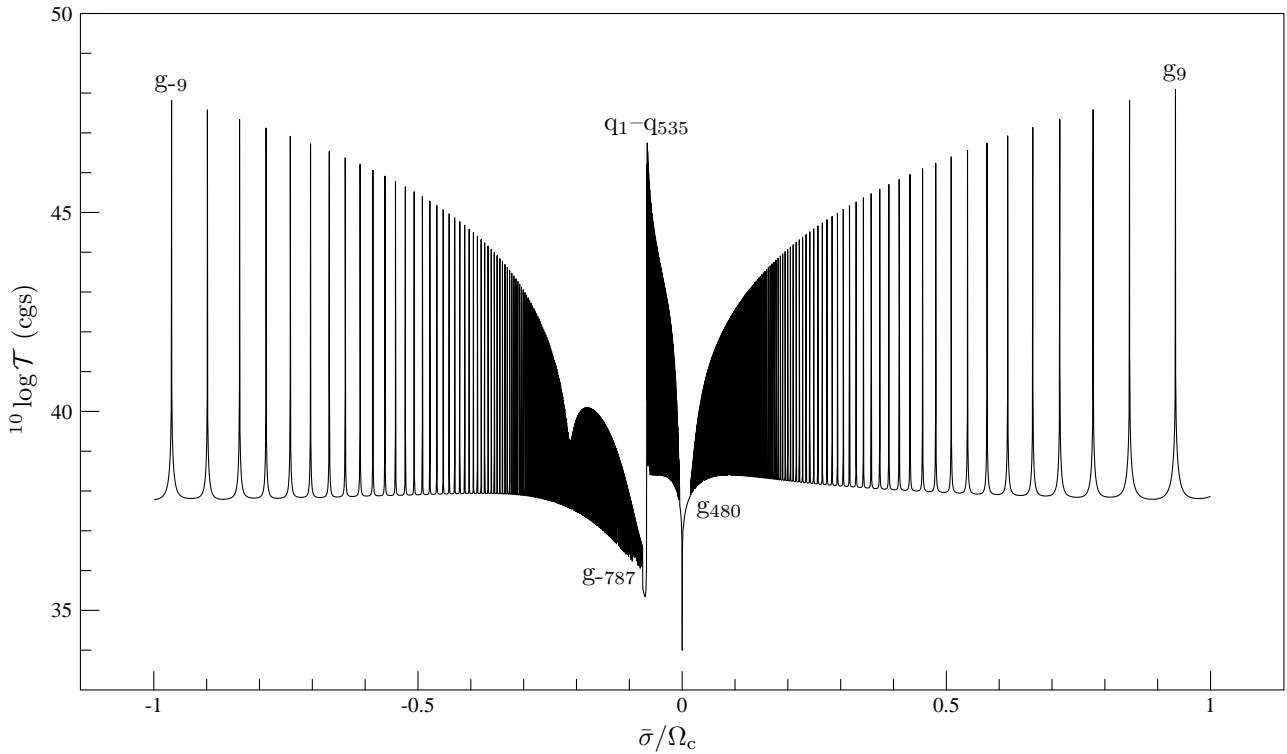}
  \caption{The torque integral versus forcing frequency for $\lambda_1>0$.
           Same as Fig.~\ref{ftork1}, but now for rotation rate 
           $\Omega_\mathrm{s}=0.2 \Omega_\mathrm{c}$.}
  \label{ftork2} 
\end{figure*}

For the prograde g-modes the resonance peaks beyond about g$_{480}$
at $\bar{\sigma} \simeq 0.03$ ($1/x\simeq 0.15$)  have been ignored,
while on the retrograde side of corotation g-modes with up to about 800 
radial nodes have been identified.
 
The off-resonant torque values follow a curved line which reflects two
main effects: for high oscillation frequencies $|\bar{\sigma}|$ the
turbulent dissipation in the convective envelope becomes less
efficient due to the mismatch of the low frequency turbulent motions
and the high frequency oscillations (Sect.~\ref{svisc}), reducing
the off-resonant torque-values.
On the other hand, for low oscillation frequencies $|\bar{\sigma}|< 0.1$ the 
peak torque-values decrease strongly by the combined effect of enhanced radiative damping 
and the rotational confinement of the g-mode response to regions close to the stellar equator.

When we increase the (uncertain) turbulent viscosity in the convective envelope 
by an order of magnitude the off-resonant torque values {\it increase} by roughly an 
order of magnitude, while the peak values at the resonances (for $|\bar{\sigma}| >0.1$ 
where radiative damping is weaker) {\it decrease} by a similar factor. 
For low forcing frequencies the tidal g-mode response has short radial wavelength, 
and the resonances are predominantly damped by radiative damping (for the adopted
turbulent viscosity law).

For $\Omega_\mathrm{s}=0.1$ the torque distribution (Fig.~\ref{ftork1}) 
shows a dip near $\bar{\sigma}\simeq -0.11$. 
The same dip occurs more prominently at about the same value
of $1/x \simeq -0.5$ for $\Omega_\mathrm{s}=0.2$ (or higher rates), see 
Fig.~\ref{ftork2}. When the convective envelope is artificially removed from 
the star the dip in the torque distribution disappears. This indicates the dip is 
related to rotation and is caused by the convective envelope. 

For frequencies $ -\Omega_\mathrm{s}/3 <\bar{\sigma}<0$ the
eigenvalues $\lambda_1$ are small (see Fig.~\ref{feigenval}a). 
These small eigenvalues are associated with the closely spaced quasi-toroidal 
q-modes which can be ordered as q$_k$, with $k$ denoting the number of radial 
nodes. Because of the long wavelength of the lower radial order q-modes these 
retrograde oscillations are weakly damped.
They couple efficiently to the $l=m=2$ dominant tidal component: 
the angular overlap integral $\zeta_1$ is relatively large, except when 
$|x|^{-1} \rightarrow 0$.
In Fig.~\ref{ftork1} and \ref{ftork2} the $\simeq 530$ 
identified q-mode resonances are not individually resolved and produce 
the black peak situated just left of corotation ($\bar{\sigma}=0$).
Although the damping of especially the low radial order (with small $\lambda$)
oscillations of this type may be underestimated by our simple treatment, 
the several hundred relatively weakly damped resonances with q-modes provide a 
significant tidal effect by which a rapidly spinning solar-type star can be spun 
down efficiently.

\subsection{Torque spectra for $\lambda_1<0$ (inertial modes) }
\begin{figure}[htbp]
  \includegraphics{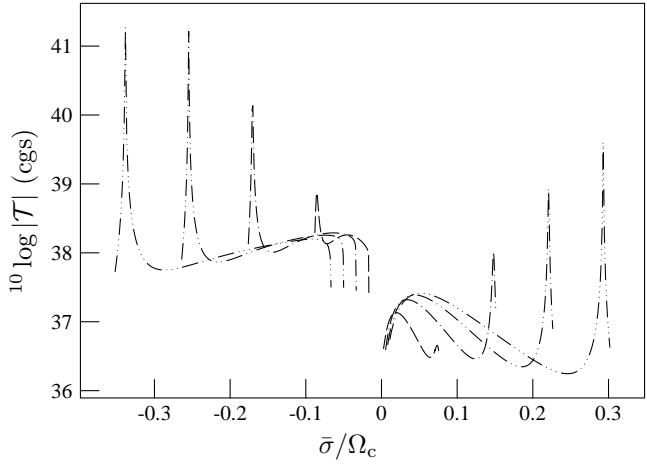}
  \caption{The tidal torque integral versus forcing frequency for $\lambda_1<0$,
    (i.e. due to excitation of inertial modes in the convective envelope) for
    several stellar rotation speeds: $\Omega_\mathrm{s}=$ 0.05,
    0.10, 0.15 and 0.20 in units of $\Omega_\mathrm{c}$. The dashed
    lines correspond to $\Omega_\mathrm{s}$=0.05, the dot-dashed
    lines to 0.10 etc. The companion mass is  1 M$_{\odot}$, while the 
    ratio $D/R_\mathrm{s}=4$ was again fixed.}
  \label{ftork3} 
\end{figure}

For negative values of the eigenvalue $\lambda_1$ the corresponding tidal oscillations 
are restricted to the convective envelope (in contrast to all other oscillation modes 
considered here, which are all limited to the radiative stellar core) and are associated 
with a spectrum of rotationally governed inertial modes. These oscillation modes with 
frequencies limited to the inertial range ($|\bar{\sigma}|< 2 \Omega_\mathrm{s}$)
produce an extended, relatively high torque level including a strong resonance peak 
(broadened by the turbulent viscosity) in the frequency range where the torque 
due to retrograde g-modes falls off sharply, filling the gap left of the q-modes: 
compare  Figs.~\ref{ftork3} and \ref{ftork1}. 
The height of the inertial torque peak increases with the stellar
rotation rate $\Omega_\mathrm{s}$ and decreases when the turbulent viscosity
is increased. Outside the strong resonance peak the torque-values increase with 
increasing viscosity. For the low-frequencies in the inertial range 
the value of the reduction factor in the expression for the turbulent viscosity 
coefficient Eq.~(\ref{eqvisc}) is practically equal to unity. The relatively 
strong inertial response can be an important driving factor for the establishment 
of resonance locking
\citepalias{1999A&A...341..842W,1999A&A...350..129W,2001A&A...366..840W}.

\section{Conclusions}
We have calculated the tidal torque on a rotating 1 $M_{\odot}$ star
due to an orbiting companion, using the traditional approximation.
We have been able to identify both g-mode and quasi-toroidal q-mode
resonances with up to $\simeq$ 1000 radial nodes in the more
evolved main sequence models by using 5000 meshpoints for the radial
grid and calculated the corresponding tidal torque spectra for stellar
models up to the end of core hydrogen burning.
The tidal torques calculated in this paper are based on a standard local mixing
length approximation for turbulent convection in the convective envelope of
solar-type stars, with a simple prescription for reduced viscous damping at high 
forcing frequencies. The resonances with low radial order tidal oscillations are
damped predominantly by the turbulent viscosity in the lower convective envelope, 
and the corresponding peak values for the tidal torque are inversely proportional
to the adopted coefficient of turbulent viscosity. Radiative diffusion in the radial
direction is also contributing to the tidal torque (strong radiative diffusion 
dominates the damping of resonances with high radial order g-modes at low forcing 
frequencies which results in weak tidal torques).

For low retrograde forcing frequencies we find a relatively
strong tidal response in the convective envelope for angular
eigenvalues $\lambda_1<0$, which corresponds to turbulent dissipation
of tidally excited  inertial modes. This  spectrum of inertial modes
may provide an efficient driving mechanism for tidal resonance locking and may
generate an enhanced tidal evolution in low-mass binary systems. 

For still smaller $|\bar{\sigma}|$ resonances with the retrograde quasi-toroidal q-modes 
(propagating in the radiative core)  become a prominent feature in the torque spectrum. 
This q-mode spectrum consists of several hundred strong, closely spaced resonances. 
When the binary parameters and stellar rotation rate are such that a low orbital harmonic 
falls in the frequency regime of the q-modes, enhanced tidal evolution must occur when
the harmonic evolves through the many closely spaced q-mode resonances. 

The effects mentioned above are related to rotation of the binary components and 
have been ignored in previous studies of tidal effects in low-mass binary systems. 
In a forthcoming paper we will apply the above results to study the tidal evolution 
of low-mass eccentric binary systems.

\bibliographystyle{astron}
\bibliography{aamnem99,bibdata}

\end{document}